\documentclass[11pt,a4paper,reqno]{amsart}%
\usepackage{amsmath,amsthm,amsfonts,amssymb,amsxtra,appendix,bookmark,dsfont,bm,color}
\usepackage[shortlabels]{enumitem}

\theoremstyle{plain}

\theoremstyle{definition}

\newcounter{step} 

\usepackage{etoolbox}
\AtBeginEnvironment{proof}{\setcounter{step}{0}}

\def\bq{\begin{eqnarray}}
\def\eq{\end{eqnarray}}
\def\bqq{\begin{align*}}
\def\eqq{\end{align*}}

\def\R {\mathbb{R}}

\def\cE {\mathcal{E}}

\def\R {\mathbb{R}}

\def\d{{\, \rm d}}
\def\Per{{\rm Per}}
\def\div{{\rm div}}
\newcommand{\Sph}{\mathbb{S}}

\title[Nonexistence in the liquid drop model]{Nonexistence of large nuclei\\  in the liquid drop model}

\author[R.L. Frank]{Rupert L. Frank}
\address{R. L. Frank, Mathematics 253-37, Caltech, Pasadena, CA 91125, USA} 
\email{rlfrank@caltech.edu}

\author[R. Killip]{Rowan Killip}
\address{R. Killip, UCLA Mathematics Department, Box 951555, Los Angeles, CA 90095-1555, USA}
\email{killip@math.ucla.edu}

\author[P.T. Nam]{Phan Th\`anh Nam}
\address{P. T. Nam, Institute of Science and Technology Austria, Am Campus 1, 3400 Klosterneuburg, Austria} 
\email{pnam@ist.ac.at}

\begin{document}

\begin{abstract}
We give a simplified proof of the nonexistence of large nuclei  in the liquid drop model and provide an explicit bound. Our bound is within a factor of 2.3 of the conjectured value and seems to be the first quantitative result.  
\end{abstract}

\date{April 14, 2016}

\makeatletter{\renewcommand*{\@makefnmark}{}
\footnotetext{\copyright\, 2016 by the authors. This paper may be reproduced, in its entirety, for non-commercial purposes.}\makeatother}

\maketitle


We consider the minimization problem
$$
E(A)=\inf\{\cE(\Omega)\!:\,  |\Omega|=A\}
$$
over all measurable set $\Omega\subset \R^3$ with the energy functional
$$ \cE[\Omega]=\Per\, \Omega + \frac{1}{2}\iint_{\Omega\times \Omega}\frac{\d x \d y}{|x-y|} \,.$$
Here $\Omega$ can be interpreted as a nucleus in the liquid drop model with density 1, and hence the volume $|\Omega|=A$ is the number of nucleons
(protons and neutrons) in the nucleus. Mathematically, $A$ is not necessarily an integer. The perimeter $\Per\, \Omega$  is taken in the sense of De Giorgi, namely
$$ \Per \, \Omega =\sup \left\{ \int_\Omega \div F(x) \d x \,|\, F\in C_0^1(\R^3,\R^3), |F|\le 1 \right\},$$
which boils down to the surface area of $\Omega$ when the boundary is smooth. The Coulomb term describes the proton repulsion in the nucleus, where the proton charge has been normalized appropriately. The liquid drop model goes back to the pioneering works of Gamow \cite{Gamow-30}, von Weizs\"acker \cite{Weizsacker-35} and Bohr \cite{Bohr-36} in 1930's, and recently it has gained renewed interest from many authors, see for instance  \cite{AlbChoOtt-09,ChoPel-10,ChoPel-11,CicSpa-13,LuOtt-14,KnuMur-14,BonCri-14,FraLie-15,KnuMurNov-15}.

It is well-known that among all measurable sets of a given volume, balls minimize the perimeter (by the isoperimetric inequality \cite{DeGiorgi-58}) and maximize the Coulomb self-interaction energy (by the Riesz rearrangement inequality \cite{Riesz-30}). This energy competition makes the liquid drop model highly nontrivial. It is generally assumed in the physics literature and conjectured in the mathematics literature \cite{ChoPel-11} that $E(A)$ is minimized by a ball up to 
$$ A_c= \frac{2-2^{2/3}}{2^{2/3}-1} \cdot \frac{|B|\ \Per B}{\frac{1}{2} \iint_{B\times B} |x-y|^{-1} \d x \d y }= 5 \cdot \frac{2-2^{2/3}}{2^{2/3}-1} \approx 3.518,$$
(see also \cite{FraLie-15}) and that for $A > A_c$ there is no minimizer.

The fact that there is no minimizer for large $A$ has been shown only recently in remarkable works of Kn\"upfer--Muratov \cite{KnuMur-14} and Lu--Otto \cite{LuOtt-14}. Their methods are inspired by techniques from geometric measure theory and seem to lead to rather large constants. In the present paper, we will provide a direct and simple proof of the nonexistence and give an explicit bound on the maximal size of a nucleus. Our main result is 

\bigskip

\noindent\textbf{Theorem}. \emph{If $A >8$, then $E(A)$ has no miminizer. }

\bigskip

This is within a factor of 2.3 of the conjectured value and seems to be the first quantitative result. Our proof builds on ideas in \cite{LuOtt-14,NamVan-16}, which were originally developed to deal with the nonexistence in the Thomas--Fermi--Dirac--von Weis\"acker theory.


\begin{proof}
Let $\Omega$ be a minimizer for $E(A)$ for some $A>0$. For every $\nu\in\Sph^2$ and $\ell\in\R$ we consider the plane
$$
H_{\nu,\ell} := \{ x\in\R^3 \,|\, \nu\cdot x = \ell \}
$$
and well as the halfspaces
$$
H_{\nu,\ell}^+ := \{ x\in\R^3 \,|\, \nu\cdot x > \ell \} \,,
\qquad
H_{\nu,\ell}^- := \{ x\in\R^3 \,|\, \nu\cdot x < \ell \} \,.
$$
We use the notation
$$
\Omega_{\nu,\ell}^\pm = \Omega\cap H_{\nu,\ell}^\pm \,.
$$
By minimality of $\Omega$ and subadditivity of the energy, we have for every $\nu\in\Sph^2$ and $\ell\in\R$,
$$
\cE(\Omega_{\nu,\ell}^+) + \cE(\Omega_{\nu,\ell}^-) \ge E(|\Omega_{\nu,\ell}^+|)+E(|\Omega_{\nu,\ell}^-|)  \ge E(A)= \cE(\Omega) \,, 
$$
which is the same as
$$
2 \mathcal{H}^2(\Omega \cap H_{\nu,\ell}) \ge  \iint_{H_{\nu,\ell}^+\times H_{\nu,\ell}^-} \frac{\chi_\Omega(x)\chi_\Omega(y)}{|x-y|} \d x \d y \,,
$$
where $\mathcal{H}^2$ denotes the two-dimensional Hausdorff measure and $\chi_\Omega$ is the characteristic function of $\Omega$. We integrate this inequality over $\ell\in\R$ and use the fact that
$$
\int_\R \mathcal{H}^2(\Omega \cap H_{\nu,\ell}) \d\ell = |\Omega|=A
$$
and
$$
\int_\R \chi_{\{\nu\cdot x>\ell>\nu\cdot y\}} \d\ell = \left(\nu\cdot(x-y)\right)_+
$$
to get
$$
2 A \geq \iint_{\R^3\times\R^3} \chi_\Omega(x) \frac{\left(\nu\cdot(x-y)\right)_+}{|x-y|} \chi_\Omega(y) \d x \d y \,.
$$
Finally, we average the bound with respect to $\nu\in\Sph^2$ and use the fact that, for any $a\in\R^3$,
$$
(4\pi)^{-1} \int_{\Sph^2} (\nu\cdot a)_+\d\nu = \frac{|a|}{2} \int_0^{\pi/2}\cos\theta \sin\theta\d\theta = \frac{|a|}{4}
$$
to conclude that
$$
2A \geq \frac{1}{4}\iint_{\R^3\times\R^3} \chi_\Omega(x) \chi_\Omega(y) \d x \d y  = \frac{A^2}{4} \,.
$$
Thus, $A\leq 8$, which proves the theorem.
\end{proof}

\noindent{\bf Acknowledgements.} The third author would like to thank H. Van Den Bosch for helpful
discussions. Partial support by U.S. National Science Foundation DMS-1363432 (R.L.F.), DMS-1265868 (R.K.) and Austrian Science Fund (FWF) Project Nr. P 27533-N27 (P.T.N.) is acknowledged.

\end{document}